\documentclass[11pt]{article}
\usepackage{a4,epsfig}

\def\Journal#1#2#3#4{{#1} {\bf #2}, #3 (#4)}

\def\NIMA{{\em Nucl. Instrum. Methods} A}

\def\PLB{{\em Phys. Lett.}  B}
\def\PRL{\em Phys. Rev. Lett.}
\def\PRD{{\em Phys. Rev.} D}
\def\RMP{\em Rev. Mod. Phys.}
\def\EPJ{{\em Eur. Phys. J.} A}

\def\be{\begin{equation}}
\def\ee{\end{equation}}
\def\bea{\begin{eqnarray}}
\def\eea{\end{eqnarray}}
\begin{document}
\title{Non--baryonic Dark Matter Searches}
\author{Y. Ramachers\\{\em \small University of Oxford, Department 
of Physics,} \\
{\em \small Nuclear and Particle Physics Lab., 1 Keble Road,}\\
{\em \small Oxford OX1 3RH, U.K.}\\{\em \small present address: LNGS, INFN, 
67010 Assergi, Italy}\\{\small E-mail:yorck.ramachers@lngs.infn.it}}
\date{}
\maketitle
\begin{abstract}The emphasise in this review about non--baryonic 
dark matter will be on experimental approaches to this fast evolving 
field of astroparticle physics, especially the 
direct detection method. The current status 
of experimental techniques will be reviewed and recent 
highlights as well as future plans will be introduced.
\end{abstract}

\section{Introduction}
\begin{figure}[htb]
\epsfxsize=10cm
\centerline{\epsfbox{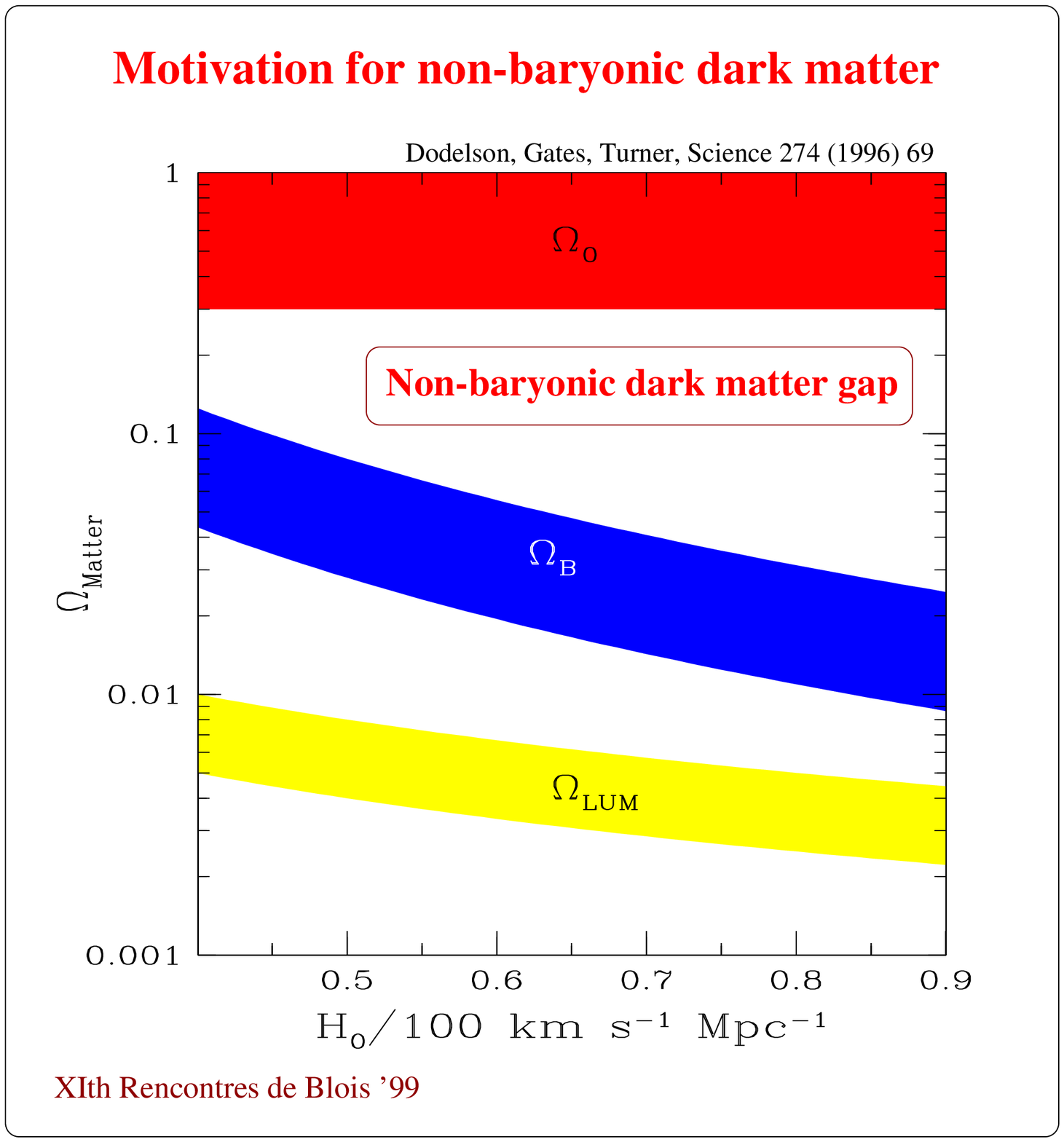}}
\caption{}
\label{motivpic}
\end{figure}

The concept of dark matter in the universe is by now well established 
(see \cite{evidence} for a collection of recent reviews) but poses 
still a remarkable problem thereby inspiring 
the creativity of astronomers and physicists. Nowadays there 
exist two separate dark matter problems, the baryonic and the non-baryonic 
dark matter problem (including a non-zero vacuum energy). That separation 
is nicely visualised in \cite{motiv} 
from which Fig.~\ref{motivpic} has been taken. The separation is founded 
on the well established constraints on the allowed amount of baryonic matter 
in the universe from primordial nucleosynthesis (see \cite{olive} 
and references therein). Every hint from 
measurements, typically on large astronomical scales, for a mass abundance 
above the allowed one is actually an evidence for non-baryonic dark matter. 
\par
As more refined measurements from experimental cosmology are collected, like 
from distant supernova Ia searches, large scale flows or cosmic microwave 
background \cite{proc}, the question of existence of dark matter 
shifts rather to 
the question of abundance and nature of dark matter, especially for 
its non-baryonic part. That part in particular is one point where 
particle physics and astrophysics merged to form the field of astroparticle 
physics. Plenty of candidates for non-baryonic dark matter have been proposed. 
Here it should be noted that the accumulating evidences for the existence 
of a finite vacuum energy, like a cosmological constant, do not render 
a non-baryonic matter contribution unnecessary. In fact, the 
numerous candidates which 
are classified as cold dark matter (non-relativistic thermal or non-thermal 
relics from the early universe) are still necessary ingredients for 
structure formation in the universe \cite{proc}. 
\par
The favourite particle candidates for non-baryonic dark matter in terms 
of experiments aiming at their detection are the axion and the neutralino. 
Since there exist extensive reviews about particle candidates and in 
particular about the axion and the neutralino, an introduction of these 
main candidates is considered to be beyond the scope of this review 
(see e.g. \cite{particle}). They can also be classified as WIMPs 
(weakly interacting massive particles) together with less prominent 
candidates (axinos, gravitinos, etc.). In fact, from the experimental 
point of view the term WIMP summarises almost all necessary requirements for 
a dark matter particle candidate. Any neutral, massive (between a few GeV and 
a few hundred TeV for thermal relics) and weakly interacting particle can 
represent a good candidate. Experiments aiming at the detection of  
particles with properties as above are described below \footnote{That 
excludes the specialised experiments aiming at the detection of axions. 
For a collection of recent references on this topic, see \cite{axion}.}
(compare also the recent review \cite{morales}).
\section{Detection concepts}
The experiments to reveal the nature and abundance of 
particle dark matter can be divided in two conceptually different 
approaches, direct and indirect detection. The physics underlying the 
direct detection technique is the elastic scattering of a WIMP with a 
nucleus of the detector. Therefore, the main observable is the deposited 
energy by the recoiling nucleus. For indirect detection the WIMPs first 
have to be accumulated in a gravitational potential to increase their 
density and therefore their annihilation rate. The annihilation products, 
high energy neutrinos, are then detected via their conversion to muons. 
Hence the signatures are muons originating from the centre of the earth 
or the sun, so called upward-going muons from these well defined sources. 
\par 
Additional observables for the direct detection technique which can serve 
to increase the signal to noise ratio are either recoil-specific or 
WIMP-specific. The recoil-specific observables are e.g. pulse-shapes or the 
partitioning of energy release into ionization and phonon signals or 
scintillation and phonon signals (see Fig.~\ref{experimpic} for a summary 
of applied techniques to reduce background). 
\par
WIMP-specific observables result from the assumption of the 
existence of a WIMP halo around our galaxy as motivated from galaxy 
rotation curves \cite{rot}. Such a halo would yield 
kinematic signatures for WIMP-detection. The movement of the sun through 
the halo (excluding a conspiracy of a co-rotating halo) induces an 
annual modulation of WIMP-rates in the detector because the earth velocity 
would add to the mean kinetic energy of WIMPs impinging on the detector 
in summer and subtract in winter \cite{mods}. The asymmetry of the WIMP-wind 
itself would also induce a diurnal modulation for a recoil-direction sensitive 
detector \cite{spergel}. 
\par 
In order to rank the various background suppression mechanisms for 
direct detection one has to keep in mind that the powerful background 
discrimination via recoil-specific observables (factors of 100 and more 
have been reported, see e.g. \cite{cdmsref,cresstref}) is systematically 
limited by neutron elastic scattering processes since these also 
produce nuclear recoils. The WIMP-specific observables are limited as well. 
First of all, the distribution function of WIMPs in the halo, see below, 
is unknown. Second, the annual modulation effect is small, of the order of 
a few percent. A recoil-direction sensitive detector would exploit the 
much larger diurnal modulations of the order of tens of percent modulations 
but detecting the tiny tracks from nuclear recoils is a formidable 
experimental task, see the DRIFT proposal below. 
\par
A comparison of indirect and direct detection methods has been worked 
out either model-independent or for a specific candidate particle 
\cite{tao,halzen,sadoulet}.
A general feature of such a comparison is that indirect searches are more 
sensitive for large WIMP-masses and spin-dependent interactions (see below)
than direct searches (see also \cite{price}). Therefore these two approaches 
are complementary. Additionally, in case both techniques would give a 
consistent signal it would be possible to obtain in principle not only 
the approximate mass and elastic scattering cross section but also 
the annihilation cross section \cite{rich}. For more details about the 
indirect detection method, I refer to \cite{price} and references therein. 
Note that in case of the neutralino as the dark matter particle candidate 
another complementarity between direct detection and accelerator 
experiments has been shown \cite{baer} (for a discussion, see \cite{yr2}). 
\par
Now for the direct detection technique, it is worth summarising the 
main experimental requirements.
\begin{itemize}
\item Energy threshold: As low as possible due to the quasi-exponentially 
decreasing signal shape as function of recoil energy. The relevant 
energy region is generally below 100 keV.
\item Target mass: As high as possible due to the low cross section for 
WIMP-nucleus elastic scattering. Direct WIMP detection means rare event 
search, i.e. the already tested rates are of the order of one count/(day kg 
keV).
\item Background: As low as possible, especially the omnipresent neutron 
flux due to the production of nuclear recoils, simulating WIMP events.
\end{itemize}
The exact dependencies of the direct detection technique on physical 
parameters can be extracted from eq.\ref{eqn1}: 
\be{}\label{eqn1}
\frac{d\,R}{d\,Q}=N_{T}\,\frac{\rho_{0}\sigma_{0}}{2\,m_{W}\,\mu^{2}}
\,F^{2}(Q)\,\int^{v_{\rm max}}_{v_{\rm min}}\, 
\frac{f(v)}{v}\,dv\quad{};\quad{}
v_{\rm min}=\sqrt{\frac{E_{thr}\,m_{N}}{2\,\mu^{2}}}
\ee{}
where dR/dQ is the measured quantity, the energy spectrum in rates over 
unit energy and unit detector mass. The other parameters can be classified 
as either completely 
unknown, like properties of the unknown WIMP, mass $m_{W}$ and 
elastic scattering cross section $\sigma_{0}$, or estimated input from 
astrophysics, like the local halo density $\rho_{0}$ (0.3-0.7 GeV/cc), 
escape velocity 
$v_{\rm max}$ ($\approx{}600$ km/s) from the galactic potential and the 
WIMP-halo distribution 
function f(v), often approximated by a Maxwellian distribution (see 
\cite{lewin} and references therein). The last set 
of values represents the number of targets $N_{T}$, target nucleus mass 
$m_{N}$, reduced mass $\mu=(m_{W}m_{N})/(m_{W}+m_{N})$ and detector 
threshold $E_{thr}$. The form factor $F^{2}(Q)$ 
parametrises the loss of coherence for the WIMP-nucleus interaction as the 
WIMP-energy increases. It represents an input from nuclear physics and 
depends on the type of WIMP interaction considered, since the elastic 
scattering cross section has two distinct interaction channels, a scalar or 
spin-independent and an axial or spin-dependent channel (for details, see 
\cite{particle} and the discussion in \cite{lewin}). Hence, depending on 
the spin-properties 
of the target nuclei, the appropriate form factor has to be used.
\section{Current Experiments}
\begin{figure}[htb]
\epsfxsize=10cm
\centerline{\epsfbox{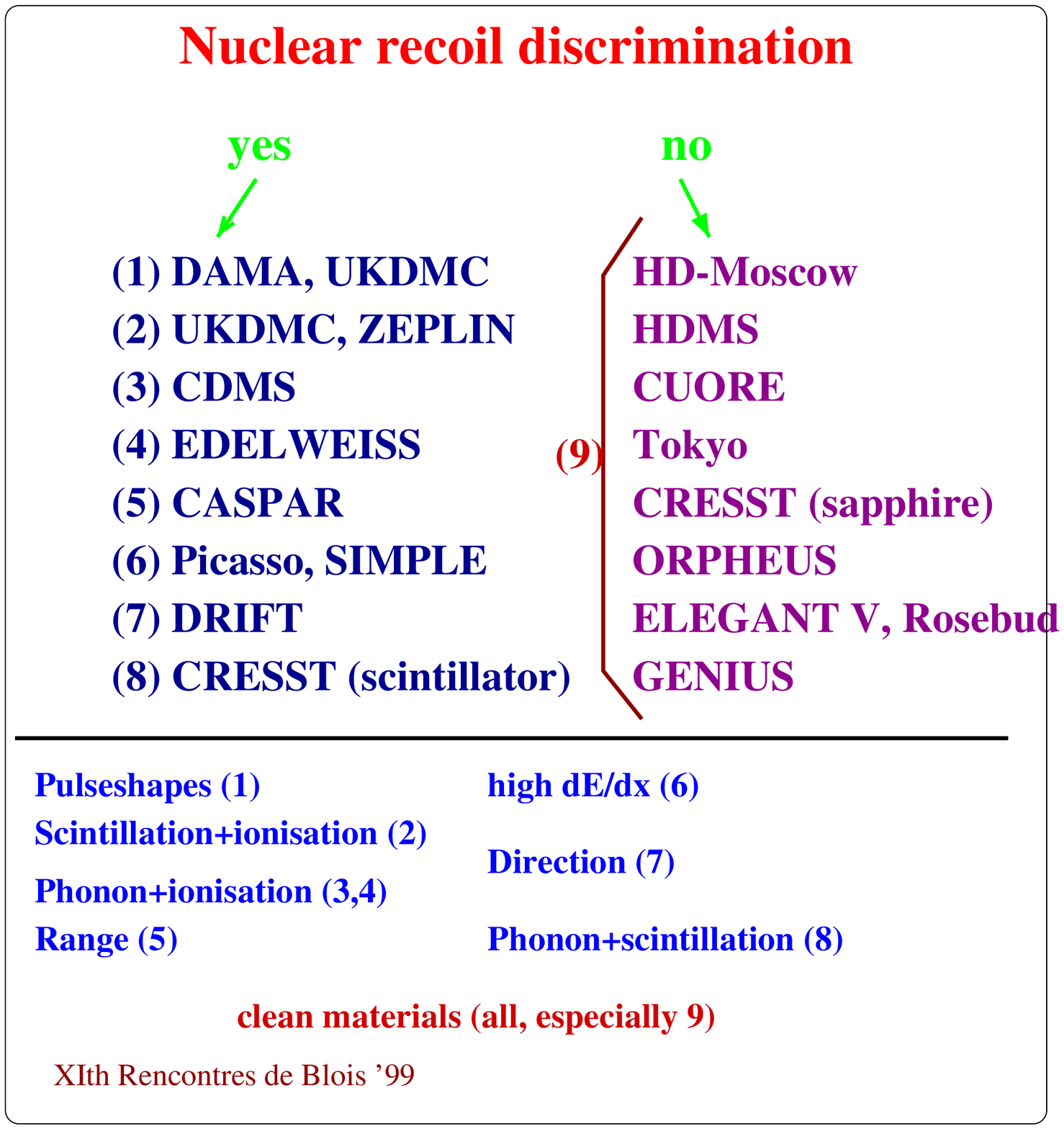}}
\caption{}
\label{experimpic}
\end{figure}

The direct detection of WIMPs became a very dynamic field of research 
during the last years. There are about 20 experiments running or being 
prepared worldwide and even more planned for the near future (for 
collections of contributions from most of these, see \cite{book1,book2}). 
Various techniques and detector systems are applied. They can be classified by 
the applied detectors \cite{morales}. Here they are classified by their 
ability to discriminate nuclear recoils to some extent, i.e. to use 
recoil-specific observables (compare Fig.~\ref{experimpic}). That separation 
of experiments might give the impression that the most advanced 
experiments will use recoil discrimination techniques. On the other hand, 
note that this particular ability also adds complexity and therefore there 
are in fact experiments which do not apply any recoil discrimination 
but, nevertheless, give competitive perspectives (see below).
\par
As shown in Fig.~\ref{experimpic}, the trick is for all no-discrimination 
detectors, like germanium semiconductor detectors (HD-Moscow \cite{hdmo}, 
HDMS \cite{hdmo}, GENIUS \cite{genius}), 
the cryogenic bolometers (CUORE \cite{cuore}, Tokyo \cite{tokyoref}, 
CRESST \cite{cresstref}, Rosebud \cite{rosebud}), the superheated 
superconducting grains detector (ORPHEUS \cite{orpheus}) or the NaI 
scintillator ELEGANT V \cite{elegant}, to use 
materials and shieldings for the setup as radio-pure as possible. Since by 
now this kind of experiments 
have already collected a large amount of experience in using clean materials 
it has been thought that their sensitivity might have reached a saturation and 
no further breakthroughs could be expected. As it turned out, this assumption 
is not true at least for germanium detectors (see the GENIUS expectation 
in the next section). 
\par
The lowest background and therefore the best limit from raw data is still 
obtained by the Heidelberg--Moscow experiment. It uses an enriched $^{76}$Ge 
detector of 2.758 kg active mass and reached a background near threshold 
of $5.7\;10^{-2}$ counts/(day kg keV). Due to the rather high 
threshold of 9 keV its limit for lower WIMP-masses can be combined with 
another germanium experiment \cite{abriola} to give a combined Ge-diode 
limit (see 
Fig.~\ref{rick}) close to the currently best limits on spin-independent 
WIMP--nucleon interactions. HDMS is a specialised WIMP-detection setup from 
the same collaboration using a germanium well-type detector as active veto 
for a small inner germanium detector. After exchanging the inner natural 
germanium crystal by a $^{73}$Ge enriched crystal the prospects are to 
improve the existing limit by about a factor of 5-10, thereby challenging 
current limits. 
\par
\begin{figure}[htb]
\epsfxsize=10cm
\centerline{\epsfbox{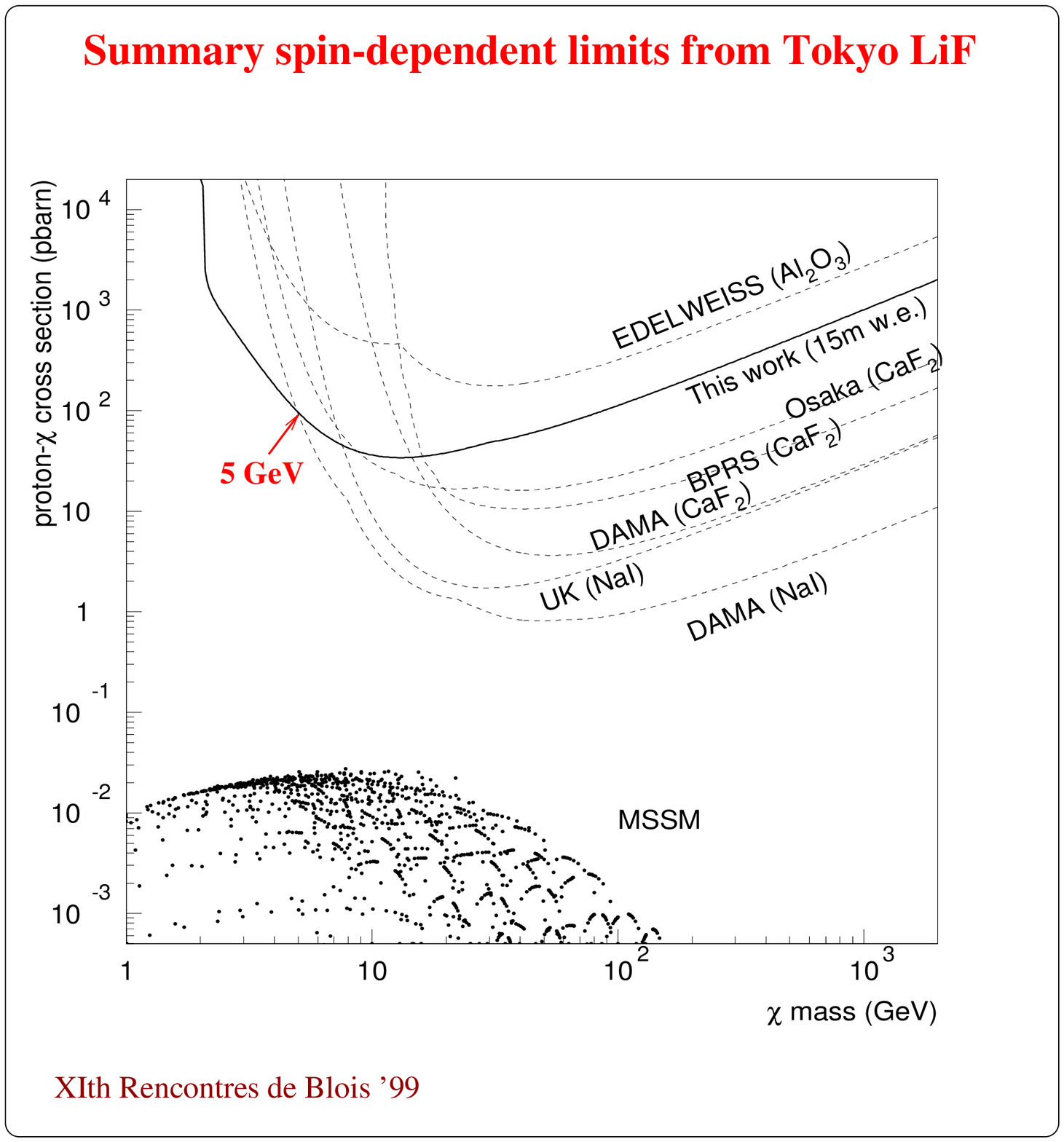}}
\caption{}
\label{tokyo}
\end{figure}

Other more complex techniques used for raw data experiments are cryogenic 
detectors. Two collaborations published first results recently, the 
LiF-crystal experiment in Tokyo \cite{tokyoref} and Rosebud \cite{rosebud} 
using sapphire (Al$_{2}$O$_{3}$) 
crystals. While the sapphire setup does not give competitive results so 
far it gives important insight into background contributions 
for cryogenic experiments in general. The LiF-experiment, although operating 
still at a shallow depth (15 m.w.e.), now improved the limit for light 
WIMP masses below 5 GeV (see Fig.~\ref{tokyo}). Due to their target 
materials both experiments are most sensitive for light WIMPs and 
the spin-dependent interaction channel. The Tokyo experiment will soon move 
to a deep underground laboratory and tries to remove identified background 
sources close to the detectors so that their estimate is to improve the 
current limit by an order of magnitude. Similar considerations have been 
put forward for Rosebud.
\par
The pulse-shape discrimination technique for NaI-scintillator detectors 
(DAMA \cite{damaref}, UKDMC \cite{ukdmc}) has been the first applied 
discrimination technique and turned 
out to be quite effective for increasing energies. Still the best limit 
on WIMP--nucleon cross sections come from the DAMA collaboration (for 
the scalar channel above 40 GeV, compare Fig.~\ref{rick}). The 
calibration of this method by 
the production of nuclear recoils from neutrons showed that these 
pulses are significantly faster than electron recoil pulses from photons 
or electrons. Recently, there emerged the feature of a class of pulses 
even faster than nuclear recoil pulses in two different experiments using 
NaI detectors (UKDMC and Modane \cite{gerbier}). These are considered 
to belong to an unknown background 
source and might even give a systematic limit of sensitivity for this 
technique. However, this effect is still under investigation and might as well 
be removed in the near future. 
\par
\begin{figure}[htb]
\epsfxsize=10cm
\centerline{\epsfbox{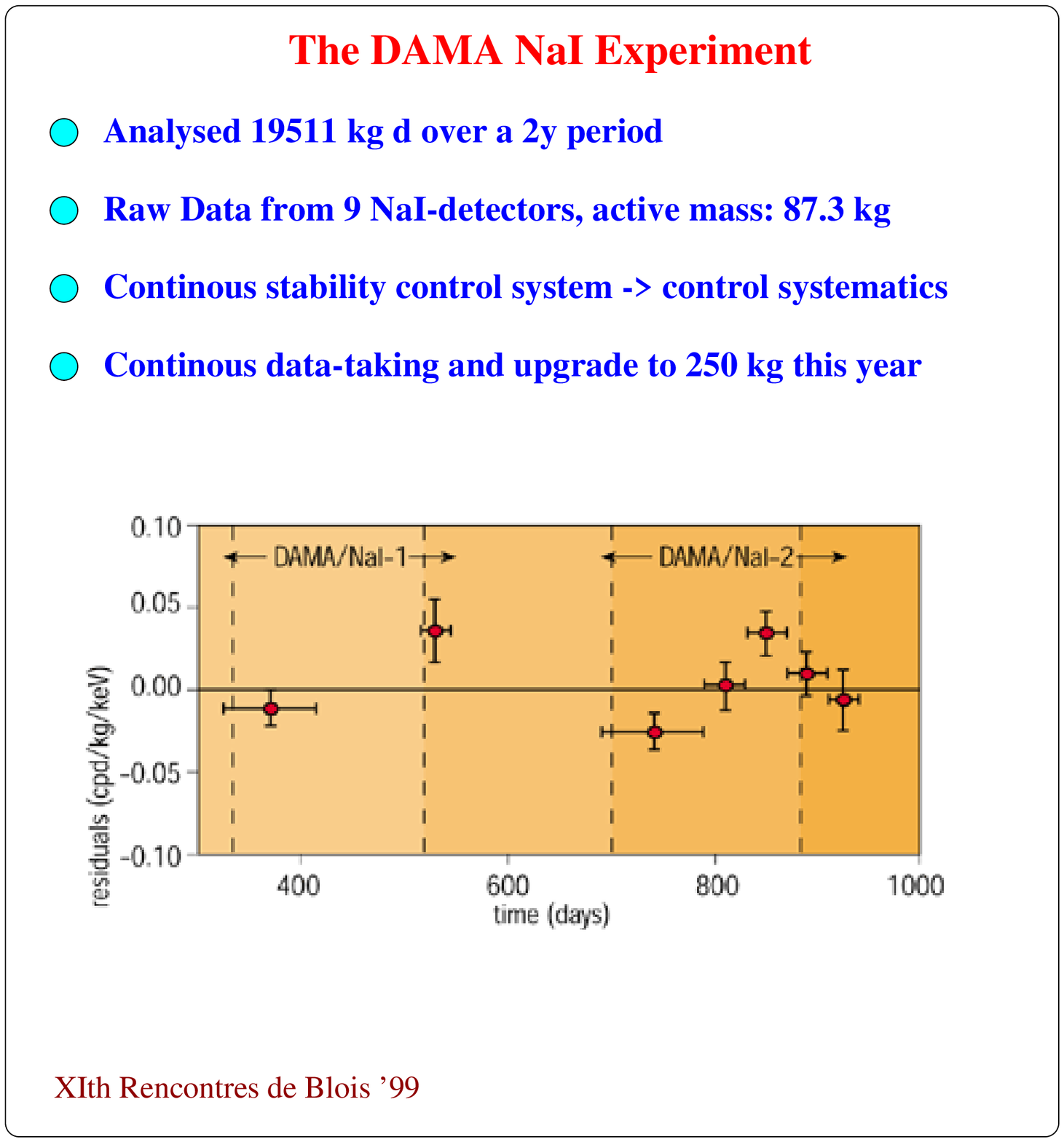}}
\caption{}
\label{dama3pic}
\end{figure}
\begin{figure}[htb]
\epsfxsize=11cm
\centerline{\epsfbox{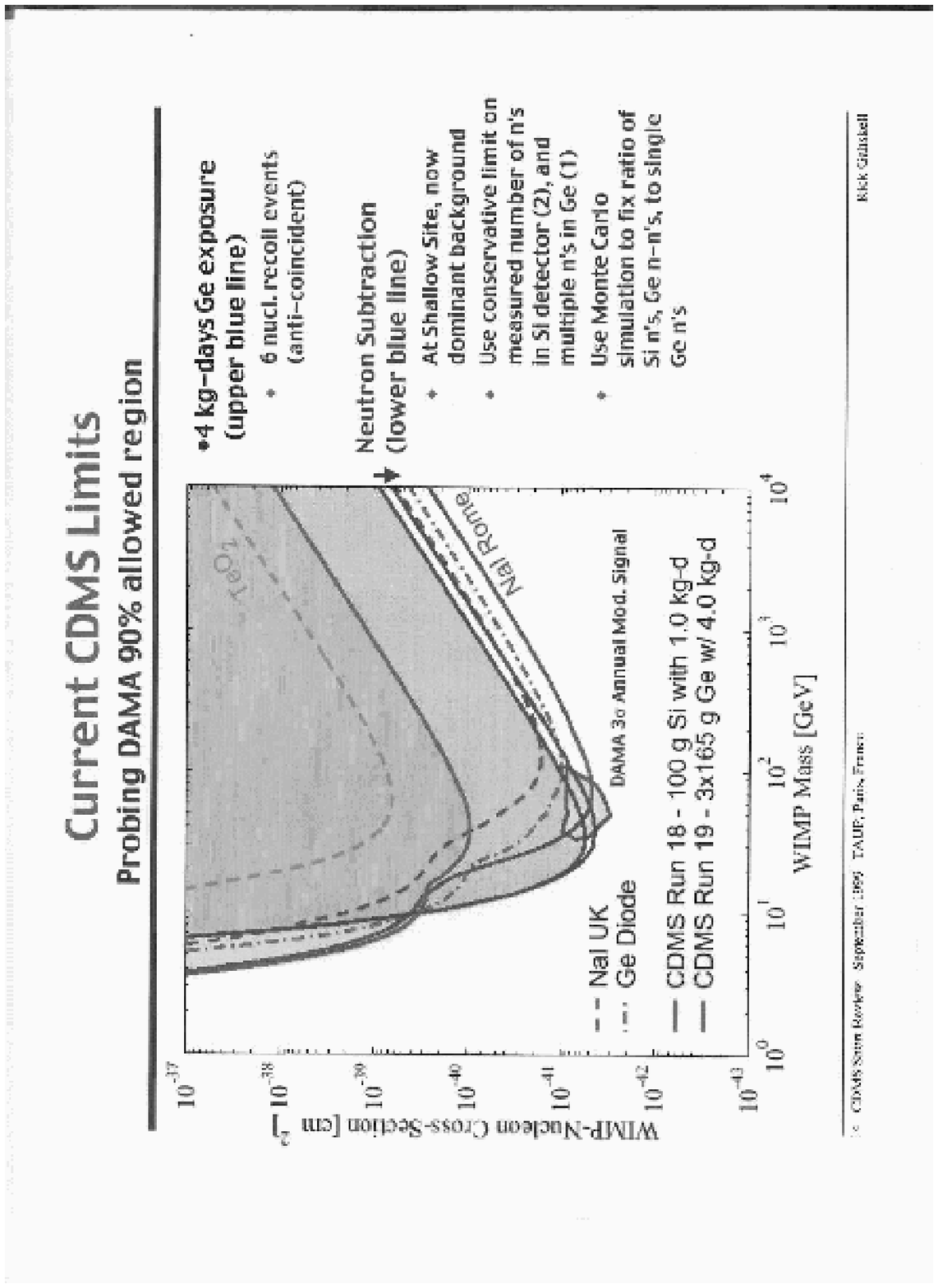}}
\caption{}
\label{rick}
\end{figure}

A highlight of NaI detectors is not only their discrimination ability 
and thereby the high sensitivity for WIMP detection but also the possibility 
to setup high target masses like the DAMA-experiment (87.3 kg active volume, 
see Fig.~\ref{dama3pic}). This puts the DAMA-experiment into the unique 
position of having the ability to even use WIMP-specific observables like 
the WIMP-signature of an annual modulation. Since that effect is just 
of the order of a few percent one has to collect a sufficient statistic 
to filter out the tiny modulation \cite{modrestriction}. In fact, this 
collaboration announced an evidence for the detection of that WIMP-signature. 
As shown in Fig.~\ref{dama3pic}, they see an annual modulation in their 
data consistent with the expectation for a WIMP at
\[
m_{W} = 59^{+17}_{-14}\;{\rm GeV}\quad{};\quad{}\xi 
\sigma_{\rm scalar}^{\rm nucleon} = 7.0^{+0.4}_{-1.2}\;10^{-6}\;{\rm pb},
\]
where $\xi{}\sigma_{\rm scalar}^{\rm nucleon}$ is the local halo density 
normalised (to 0.3 GeV/cc) WIMP--nucleon scalar cross section. The 
consistency requirements are the proper kinematic modulation (phase, 
amplitude and period), single hits in detectors and the proper signal 
shape (maximum signal in the lowest energy bin and subsequent decrease). 
However, the excitement about this evidence is accompanied by a similarly 
engaged criticism \cite{critic} inside the dark matter community. 
Fortunately, this is a matter of experiment, i.e. it will be possible 
already in the near future to test the evidence by experiments rather 
than arguments. 
\par
The first competitors of the DAMA experiment which are expected to be 
able to test their result in the near future are the cryogenic detector 
experiments CDMS and EDELWEISS \cite{edelweissref}. They use a 
combined signal readout of 
phonon signals and ionization signals from germanium (and silicon in case 
of CDMS) crystals. The clue of this kind of readout scheme is that 
in germanium crystals the ionization efficiency of nuclear recoils is 
just about 25\% (energy dependent, see \cite{efficiency} and references 
therein) compared to an electron recoil event. So far, they both 
suffered from the effect of incomplete charge collection of surface 
electron events which could mimic nuclear recoil events. Recently the 
CDMS experiment got rid of this problem (see Fig.~\ref{rick} and 
\cite{rickref}) and now they already test the DAMA result below about 
$m_{W}=40$ GeV. The EDELWEISS collaboration is expected to follow that 
development soon and release a new improved limit comparable to or even 
better than the CDMS result. 
\par
Apart from that actual status report it is worth mentioning rather mid-term 
projects which show the variety of applied detection techniques in order to 
reduce background. The CASPAR proposal \cite{caspar} uses small grains 
of CaF$_{2}$ scintillators (of the order of a few hundred nanometer 
diameter) dissolved in an organic scintillator. Calcium or fluorine 
nuclear recoils would only produce a scintillation signal from their grain 
whereas electron recoils would have a much larger range and would 
give signals from the organic scintillator as well which then could 
be discriminated. Another discriminating detector concept using ionisation 
and scintillation signal readout from liquid Xenon (or two--phase Xenon, gas 
and liquid), the ZEPLIN \cite{zeplin} project, is still in its R\&{}D--phase 
but first tests are very promising. 
\par
The superheated droplet detectors PICASSO \cite{picasso} and 
SIMPLE \cite{simple} also use the specific 
energy loss of nuclear recoils to discriminate against minimum ionising 
particles. They use a well known technique for neutron dosimetry, namely 
droplets of a slightly superheated refrigerant liquid embedded in a gel. 
The droplets would then act like tiny bubble chambers, exploding due to 
a nuclear recoil event. By tuning the relevant parameters, pressure and 
temperature, the bubbles can be made insensitive to nuclear radiation so 
that practically only recoils from fission processes and neutrons remain 
background sources. Both experiments are currently 
build up and first results from PICASSO and SIMPLE 
have been reported recently.
\section{Future Plans}
Although research in the field of direct detection evolves rapidly and 
more exciting results can be expected for the near future as mentioned 
above, there are three proposed detection concepts which shall be reported 
separately in order to point out their exceptional prospects
\footnote{Admittedly, due to personal preferences.}. 
\par
\begin{figure}[htb]
\epsfxsize=10cm
\centerline{\epsfbox{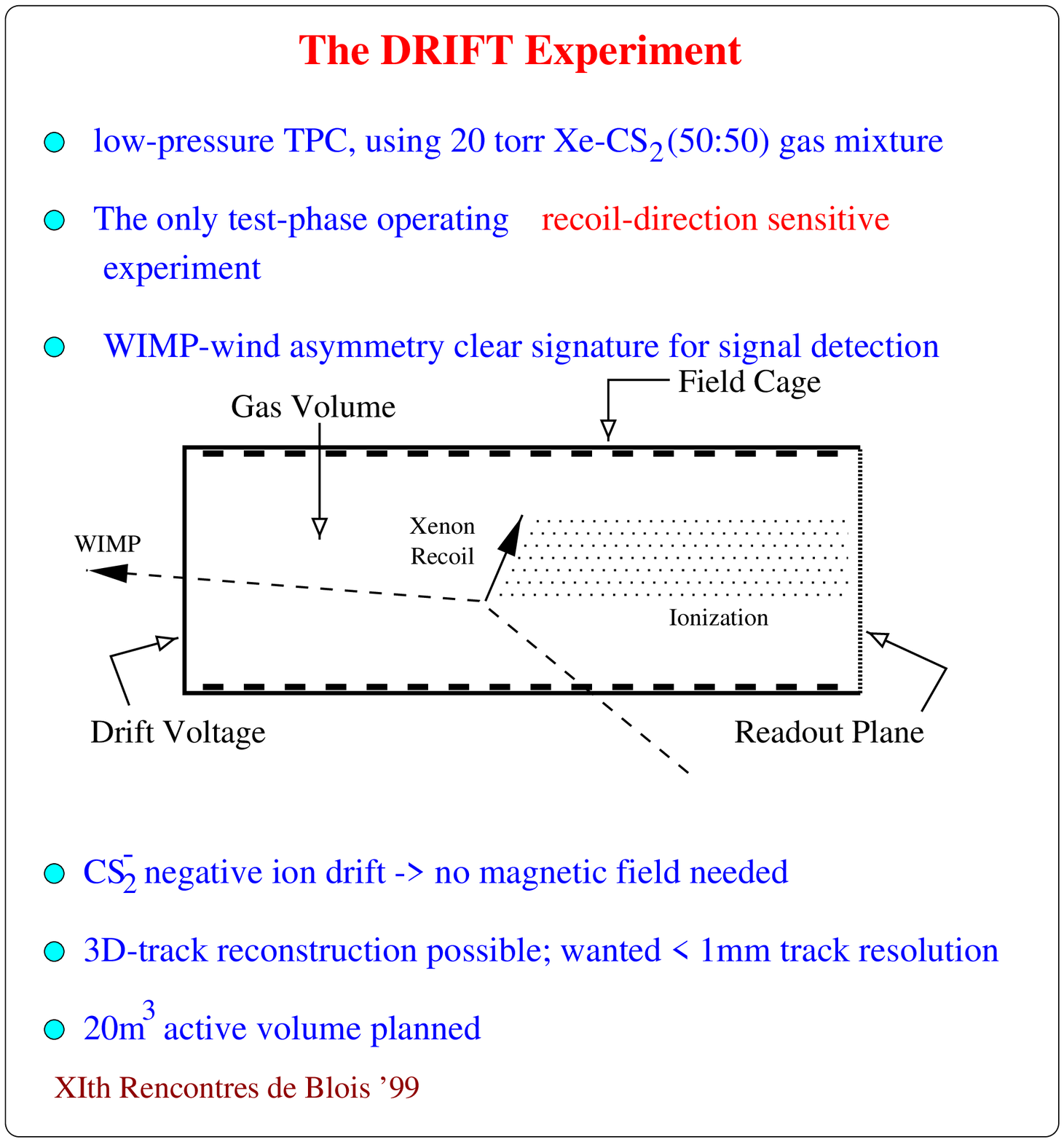}}
\caption{}
\label{drift}
\end{figure}

The DRIFT experiment (see Fig.~\ref{drift}) \cite{driftref} represents 
the only 
test-phase operating recoil-direction sensitive experiment. It consists 
of a low-pressure TPC using 20 torr Xe-CS$_{2}$ (50:50) gas mixture. The 
crucial point for such a device is that it must be able to detect reliably 
the tiny tracks from nuclear recoils, i.e. the design goal is to achieve 
a less than 1 mm track resolution. In order to setup higher target masses 
despite the low-pressure gas, the idea is to abandon the magnetic field. 
That would in principle worsen the track resolution due to enhanced diffusion. 
Therefore the detection concept has been changed in the sense that the TPC 
does not detect the drifted electrons but rather negative CS$_{2}^{-}$ ions 
with considerably reduced diffusion. The prospects for this setup are very 
encouraging due to the ability to measure the most decisive WIMP-signature, 
the diurnal modulation due to the WIMP-wind. The plan is to operate a 
20 m$^{3}$ TPC by the end of 2001.
\par
\begin{figure}[htb]
\epsfxsize=10cm
\centerline{\epsfbox{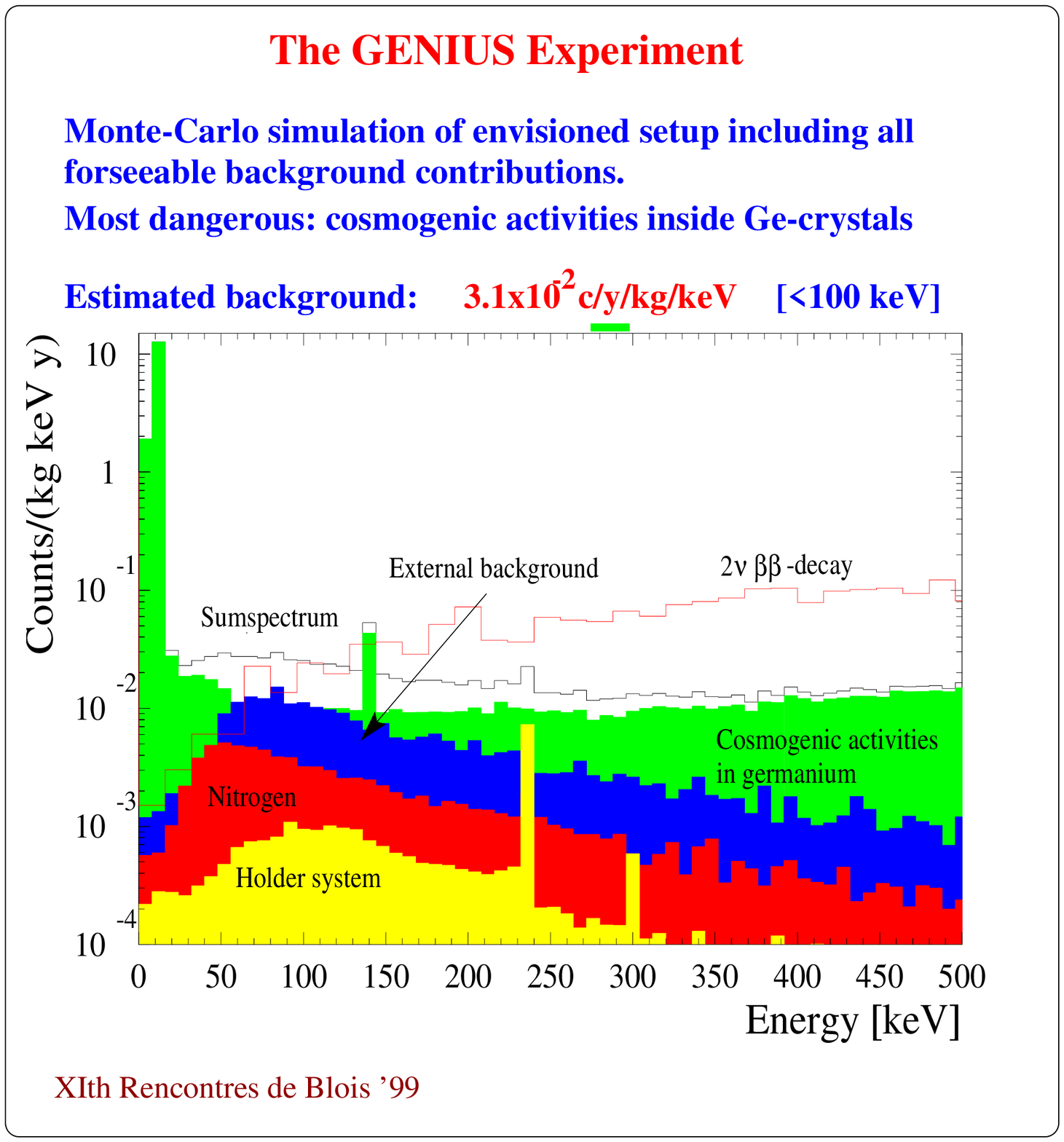}}
\caption{}
\label{geniuspic}
\end{figure}

The GENIUS proposal \cite{genius} is exceptional in the sense that it 
is a detection 
concept which works without specialised background discrimination 
procedures, i.e. it will not use nuclear recoil specific observables. 
This traditional detection method, using germanium semiconductor detectors, 
relies therefore on the utilisation of extreme radio-pure materials around 
the detectors. The idea is to remove all materials close to the crystals 
which were so far necessary to cool the detectors and instead operate 
them directly in liquid nitrogen which has been measured to be a very pure 
environment. The necessity to use that liquid nitrogen as shielding material 
scales the setup to a rather large size (dewar of 12 m diameter and height). 
On the other hand, that also admits to operate a large target mass (100 kg 
of natural germanium is planned for the first stage) in a common environment. 
The technical studies of operating 'naked' germanium crystals in liquid 
nitrogen have already been performed successfully. An estimation of 
the expected sensitivity, i.e. the background expected, can be seen in 
Fig.~\ref{geniuspic}. Most dangerous appears the cosmic activation of the 
crystals while produced and transported on the earth surface due to 
spallation reactions with cosmic rays (hadronic component). The cited 
background expectation of $3.1\;10^{-2}$ counts/(y kg keV) below 
100 keV would result in a WIMP-sensitivity more than 3 orders of 
magnitude below current best limits which definitely is an encouraging 
prospect. 
\par
\begin{figure}[htb]
\epsfxsize=10cm
\centerline{\epsfbox{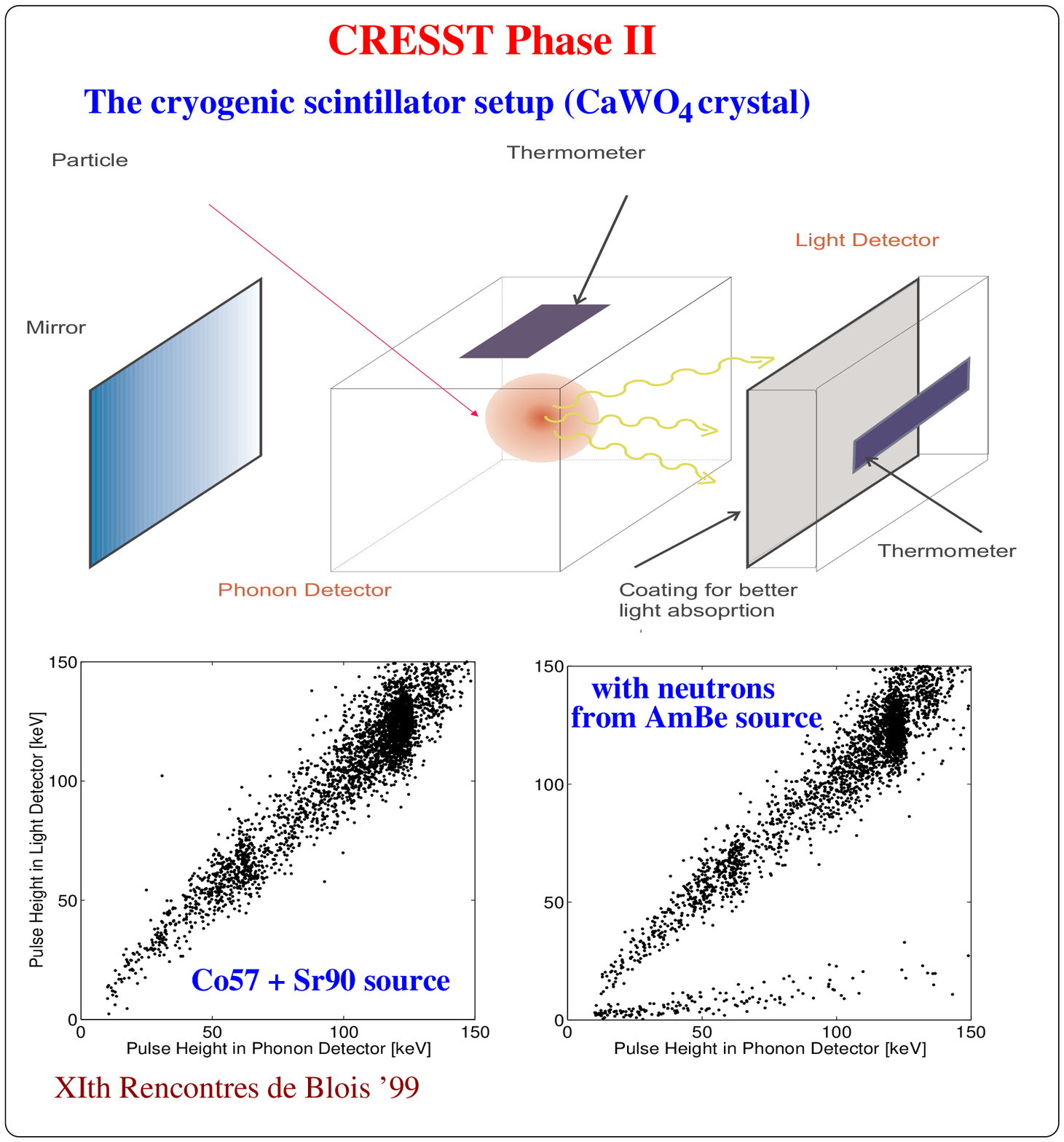}}
\caption{}
\label{cresst}
\end{figure}

The CRESST phase II concept \cite{cresstref,cresst2} 
consists of a combined signal readout from a 
scintillating crystal cooled to 12 mK. The light and phonon readout yields 
a very efficient discrimination mechanism as can be see in Fig.~\ref{cresst}. 
Test measurements using a non-optimised setup in a surface laboratory already 
give background suppression factors comparable to the ionization--phonon 
readout schemes seemingly without the problem of surface effects. Several 
scintillating crystals have been tested (BaF$_{2}$, BGO, PbWO$_{4}$ and 
CaWO$_{4}$) and their light yields at cryogenic temperatures 
measured. The operation of the scintillator as a cryogenic calorimeter 
poses special problems for the light detection since no light guide touching 
the crystal surface (matching refractive indices) is allowed since 
that would distort the phonon signal. For light detection a second 
calorimeter, 
a thin sapphire crystal coated with silicon to improve light absorption, 
is placed next to one surface of the scintillator and the other surfaces 
are surrounded by mirrors (compare Fig.~\ref{cresst}) \footnote{New 
measurements use diffuse Teflon reflectors instead, resulting in a factor two 
per unit area more efficient light yield \cite{cresst2}.}. Apart from the 
discrimination abilities of the detector concept there is an 
additional advantage. Several target materials or scintillators can be 
used in such kind of setup which would result in an unique handle 
not only on the discrimination of the ambient neutron background, so 
far considered to be the ultimate limiting factor, but also on the 
extraction of the WIMP signal due to its target mass dependence (see 
eq.\ref{eqn1}). Moreover, already in the existing cryogenic setup inside 
the Gran Sasso underground laboratory there is enough space to house 
some tens of kilogramms of active mass, rendering a modulation signature 
search possible. A first scintillation detector made from CaWO$_{4}$ is 
expected to be mounted underground already at the beginning of 2000.
\section{Conclusion}
Non-baryonic dark matter is by now a well motivated concept from astronomy 
in the framework of a universe model containing cold dark matter. Several 
independent measurements from experimental cosmology indicate the necessity 
of a matter content above the allowed baryonic matter from primordial 
nucleosynthesis. In addition, particle physics offers attractive candidates 
for cold dark matter classified as WIMPs and initially motivated independent 
from cosmological reasoning (especially the neutralino as necessary 
ingredient of supersymmetric theories). 
\par
WIMP searches in the form of direct and indirect detection experiments are 
a very active field of research also because of the attractive 
interdisciplinarity between astro- particle- and nuclear physics. A large 
variety of direct detection experiments, on which this review focused, 
currently produce results or will start in the near future. In addition, 
the first WIMP-detection evidence has been announced and will soon be tested 
by independent experiments. The benefit from this kind of research is twofold 
and worth to be reminded. One would learn about the supposed major 
part of matter in the universe and about beyond standard model physics 
by detection of non-baryonic dark matter.
\section*{Acknowledgement}
The author thanks the following researchers for providing informations about 
their experiments and valuable comments: R. Bernabei, G. Chardin, D.B. Cline, 
J. Collar, S. Cooper, H. Ejiri, M. Di Marco, J. Hellmig, H.V. 
Klapdor-Kleingrothaus, M. Lehner, L. Lessard, M. Minowa, K. Pretzl, B. 
Sadoulet, W. Seidel, N. Smith, N.J.C. Spooner, D. Tovey and HanGuo Wang.

\newpage
\section*{Figure captions}
Figure 1: Shown is a summary of astronomical results for 
the mean matter 
density in the universe combined with a conservative estimate from 
primordial nucleosynthesis as a function of the Hubble constant. The dark 
dividing line titled $\Omega_{B}$ in the middle gives the allowed amount 
of baryonic matter in the universe (the lower band gives the amount of 
visible matter). The gap between $\Omega_{B}$ and the observed 
matter density on large scales (summarised as everything above 
$\Omega_{0}=0.3$) represents the non-baryonic dark matter motivation 
(from \protect{\cite{motiv}}).
\par
Figure 2: Summary of existing and planned direct detection experiments, 
classified according to their ability to discriminate nuclear recoils. The 
numbers to the left indicate the applied background reduction technique which 
is given in the legend below. Note the variety of methods which gives a hint 
on the diverse experimental techniques and detector concepts involved 
in this fast evolving field of research.
\par
Figure 3: Collection of spin-dependent cross section 
limits for several 
direct detection experiments as function of the WIMP-mass (from 
\protect{\cite{tokyoref}}). 
Note the improved limit below 5 GeV and the 
large distance of limits to expectations (for neutralinos) for this 
particular interaction channel.
\par
Figure 4: Short summary of the most intriguing result from the DAMA NaI 
experiment (from \protect{\cite{damaref}}). Note that the upgrade to 
250 kg has been approved.
\par
Figure 5: Summary of current WIMP-nucleon cross section limits for 
spin-independent interactions from the CDMS collaboration 
\protect{\cite{rickref}}. 
The best 
limit for WIMP-masses above 40 GeV stem from the DAMA collaboration, below 
CDMS now tests the DAMA evidence contour. The combined Ge-diode limit is 
shown as dash-dotted line and dashed the UKDMC NaI result 
(limited by the unknown fast pulse shape component, see text).
\par
Figure 6: Summary of the DRIFT experiment and schematic 
view of their TPC 
setup \protect{\cite{driftref}} (for details, see text). 
\par
Figure 7: Shown are the background expectations 
for GENIUS according 
to Monte-Carlo simulations \protect{\cite{genius}}. Contributions 
from liquid nitrogen, the holder 
system, external background and cosmic activation have been included. Also 
shown is the sumspectrum and the contribution from the two-neutrino 
double beta decay of $^{76}$Ge in the natural germanium crystals.
\par
Figure 8: CRESST phase II setup and test measurement results 
\protect{\cite{cresstref}} (for details, see text).
\newpage
\par
\par 
\par
\par
\par
\par
\par
\par
\end{document}